# High-temperature operation of a silicon qubit

Keiji Ono[1*], Takahiro Mori[2], and Satoshi Moriyama[3]

[1] Advanced Device Laboratory, RIKEN, 2-1 Hirosawa, Wako, Saitama 351-0198, Japan

[2] Nanoelectronics Research Institute (NeRI), National Institute of Advanced Industrial Science and Technology (AIST), Central 2, 1-1-1 Umezono, Tsukuba, Ibaraki 305-8568, Japan

[3] International Center for Materials Nanoarchitectonics (WPI–MANA), National Institute for Materials Science (NIMS), 1-1 Namiki, Tsukuba, Ibaraki 305-0044, Japan

*k-ono@riken.jp

## Abstract

This study alleviates the low operating temperature constraint of Si qubits. A qubit is a key element for quantum sensors, memories, and computers. Electron spin in Si is a promising qubit, as it allows both long coherence times and potential compatibility with current silicon technology. Si qubits have been implemented using gate-defined quantum dots or shallow impurities. However, operation of Si qubits has been restricted to milli-Kelvin temperatures, thus limiting the application of the quantum technology. In this study, we addressed a single deep impurity, having strong electron confinement of up to 0.3 eV, using single-electron tunnelling transport. We also achieved qubit operation at 5-10 K through a spin-blockade effect based on the tunnelling transport via two impurities. The deep impurity was implemented by tunnel field-effect transistors (TFETs) instead of conventional FETs. With further improvement in fabrication and controllability, this work presents the possibility of operating silicon spin qubits at elevated temperatures.

## MAIN TEXT

### Introduction

Silicon technology is a core technology supporting our modern advanced information society. Quantum technology is an emerging technology that could become a core next-generation technology by ensuring its compatibility with silicon technology. Therefore, it is important to develop quantum technology based on silicon technology. In particular, it would be extremely beneficial with regard to the applicability of quantum sensing and quantum authentication technologies if these technologies were integrated with mature silicon technology.

Large-scale integration and high fidelity have been significant issues in Si qubit research[1, 2, 3, 4, 5, 6, 7, 8, 9, 10]. Here, we highlight another important issue, which is temperature of operation. The operational temperature of Si qubits has received little attention. Thus, at present, most Si qubits operate in the milli-Kelvin temperature range. Obviously, high-temperature operation of Si qubits could extend their application, facilitated by the use of smaller refrigerators. Therefore, from the viewpoint of practical application, it is meaningful to explore the possibility of high-temperature operation of silicon qubit, even if their potential scalability is not immediately realised.

In this paper, we propose and demonstrate a novel Si qubit by taking advantage of an individual deep impurity embedded in Si tunnel field-effect transistors (TFETs) for electrically addressable spin qubits, to realise higher-temperature operation. Deep impurities are advantageous for qubit operation because electrons trapped in such deep levels are not excited thermally. Therefore, qubit operation can be expected even at room temperature if the level is sufficiently deep. Indeed, some reports have suggested the feasibility of room-temperature operation: the ensemble characteristics of spins bound to deep impurities have been investigated using electrically detected magnetic resonance[11], and the electron spins of dangling bond defects and neighbouring $^{29}$Si nuclear spins have been detected in MOS field-effect transistors (MOSFETs) at room temperature[12]. Numerous types of deep impurities in Si, including defects, have been identified and intensively studied, and conventional shallow impurities have been utilised in previous studies on Si qubits. Deep impurities have also been investigated in conventional Si FET devices from the perspective of their infrared responses or transistor switching stabilities[13, 14].

Quantum-dot-like transport has been achieved using shallow impurities in the channels of miniaturised MOSFETs[15, 16, 17, 18]. Specifically, this transport involved two-step tunnelling from a source to a drain by utilising the shallow impurity level. Transport occurred due to sufficient tunnel coupling between the shallow level and the source and drain electrodes. However, deep impurities in MOSFETs are not suitable for such impurity-level-mediated transport because tunnel coupling is not ensured. To realise tunnel coupling with deep impurities, it is necessary to utilise extremely short-channel MOSFETs, which cannot be achieved using the current technologies. Even if such short-channel MOSFETs were realised, the thermally excited diffusion current would superimpose the tunnelling current at the required higher operation temperatures.

Therefore, we employed TFETs in this study. TFETs are gated p-i-n diodes[19] that look like MOSFETs but have different types of sources and drains. For example, P-type TFETs have n-type sources and p-type drains. Their switching is realised via electrostatic gate control by changing the thickness of the pn junction at the source-side edge of the gate. TFETs are attractive as future building blocks for low-power-consumption large-scale integrated circuits (LSIs), as they can achieve switching more abruptly than MOSFETs. ON current enhancement was reportedly achieved by introducing deep impurities into (relatively long-channel) Si TFETs and was ascribed to deep-level-assisted resonant tunnelling in their pn junctions[20, 21, 22].

In this study, we utilised short-channel TFETs with deep impurities to realise electrical access to single deep impurity levels, and utilised their spins as qubits operable at high temperatures. Not only tunnelling transport through deep levels, but also gate tuning of such levels, is possible in short-channel TFETs with appropriately located deep impurities. Unlike MOSFETs, TFETs can enable tunnel coupling between impurities and electrodes with feasible channel lengths (sub-100 nm), even to the deepest levels in the middle of their band gaps.

Some of the devices that we fabricated exhibited clear signatures of single-dot-like characteristics even at room temperature due to the strong electron confinement to their deep impurities. Some other devices exhibited spin blockade signatures[23] due to the combination of deep and shallow levels to produce double quantum dots. In addition, under microwave driving, the weak source-to-drain current in the spin blockade region exhibited electron spin resonance, which enabled time-ensemble measurements of the spin qubits to be performed. Rabi oscillation was clearly observed with microwave pulses at 1.5 K and 5 K, and was still evident at 10 K.

## Results

**Device and measurement**

The TFET-based quantum dot devices (Fig. 1) were fabricated using manufacturing processes similar to those for conventional MOSFETs (see the supplementary information for details). We utilised 100-mm Si-on-insulator (SOI) wafers, and the source and drain, which were n-type and p-type, respectively, were formed by ion implantation (I/I) of shallow donors and acceptors. The source and drain were activated by high-temperature rapid thermal annealing. Then, additional I/I of Al and N was performed throughout the Si area consisting of the source, channel, and drain. Next, low-temperature annealing, which is known to form Al–N impurity pairs in Si, was conducted[20, 21, 22]. These additional processes were crucial in forming deep impurity levels in the TFETs. Indeed, when they were skipped, none of the TFETs exhibited quantum-dot-like transport (see the supplementary information for more detailed discussion). Finally, the MOS gate, which utilised conventional high-$k$/metal gate technology, was formed. Quantum-dot-like transport was observed in the devices with gate lengths shorter than 80 nm (see the supplementary information for details).

The devices were characterised in a cryostat at 1.5–300 K. In most cases, we measured the drain current $I_D$ while the source voltage $V_S$ and the gate voltage $V_G$ were applied. Some characterisation was performed while $I_S$ was measured and $V_D$ was applied to avoid gate leakage current. Magnetic fields were applied perpendicular to the substrate, which corresponded to the [100] direction, or parallel to the source-to-drain current, which corresponded to the [110] direction. The microwaves were applied to plates located near the devices or the back plates of the substrates. In our microwave setup, a microwave electric field was dominantly applied to the device; hence, the electron spin resonance that we discuss later is an electric-dipole spin resonance [24].

**Room-temperature single-electron transport**

The short-channel devices (≤80 nm) exhibited quantum-dot-like characteristics with large single-electron charging energies, as shown in the intensity map of the differential conductance $dI_D/dV_S$ in Fig. 2(a). The Coulomb diamond $dI_D/dV_S$ suppression patterns are typical of single-quantum-dot devices, and the single-electron charging energies estimated from the widths of these Coulomb diamonds were up to 0.3 eV.

It should be noted that the intensity map is asymmetric with respect to $V_S$. Specifically, the $dI_D/dV_S$ intensity in the positive $V_S$ range is much greater than that in the negative $V_S$ range. This asymmetry is also reflected in the $I_D$–$V_S$ curves in Fig. 2(b). These features indicate that quantum dots were located in the channels (intrinsic regions) of the TFETs. Thus, tunnel coupling between the dots and electrodes depends on $V_S$ because the width of the space charge region of a p-i-n structure depends on $V_S$. Thus, negative $V_S$ results in weak tunnel coupling. This characteristic is a notable feature of TFET-based quantum dot devices because MOSFET-based quantum dot devices exhibit less sensitivity to $V_S$ but greater sensitivity to $V_G$.

Lifting of the Coulomb blockade is observable in Fig. 2(a) around $V_G = -0.25$ V. At low temperatures, the zero-bias source-to-drain conductance $G$ shown in Fig. 2(c) exhibits sharp peaks originating from the lifting of the Coulomb blockade. Notably, the conductance peak is observable even at 300 K, indicating that the device worked as a room-temperature single-electron transistor[25, 26].

Considering the large charging energy, $V_S$-asymmetric transport, and high-temperature single-electron transport, we concluded that quantum-dot-like transport was realised with single deep impurity levels, as schematically illustrated in Fig. 2(d). We suppose that a deep impurity spatially located approximately halfway between the source and drain contributes to the transport. If a deep impurity is located near the source side, the impurity level is significantly lower than the Fermi energy of the electrodes. Similarly, if a deep impurity is located near the drain side, the impurity level is significantly higher than the Fermi energy of the electrodes. Thus, it does not contribute to the transport (i.e., is outside the transport window). Therefore, a limited number of impurities are within the transport window, and the other impurities do not contribute to the transport when $|V_G|$ is small, although numerous deep impurities existed in the channels of the devices fabricated in this study.

**Single-electron spin resonance**

In some other devices, we also observed double-dot-like characteristics attributable to two quantum dots connected in series between the source and drain. Figure 3(a) presents a $dI_S/dV_D$ map obtained from one such device (device B). The Coulomb blockade region is nearly lifted at $V_G = 0.25$ V, whereas a finite gap is evident at $V_D \sim 0.02$ V. Notably, the edge of the Coulomb diamond has a zig-zag pattern at some places, such as near $(V_D, V_G) = (0.1$ V, $0.1$ V); $I_S$ peaks, accompanied by dark blue and red in the $dI_S/dV_D$ map, are observable outside the Coulomb blockade region; the $V_G$ positions of the $I_S$ peaks weakly depend on $V_G$. These features were previously observed in single-gated double quantum dots[27]. In addition, the temperature dependence of $G$ as a function of $V_G$ exhibited two types of Coulomb blockade oscillations: multiple-dot-like oscillations at temperatures below 30 K and single-dot-like oscillations at temperatures above 40 K (see the supplementary information for details)[28]. Considering all of these features, the formation of multiple dots consisting of deep impurities with strong confinement (supposing a confinement energy of >0.2 eV, see the supplementary information) can be expected, together with at least one "satellite dot" near each deep impurity with weaker confinement (~5–20 meV).

Among the $I_S$ peaks, one very sharp peak with a full width at half-maximum (FWHM) of 0.37 mV is evident in Fig. 3(b). With the $I_S$ peak, we performed microwave irradiation and observed a typical double dot feature of the Landau-Zener-Stückelberg-Majorana (also known as photon-assisted tunnelling) interference pattern (see the inset of Fig. 3(b))[29, 30, 31]. The $I_S$ peak originated from resonant tunnelling through two quantum dots because its 0.06-meV width is much smaller than the thermal energy of the measurement temperature, 1.5 K[32]. It should be noted that the peak width was estimated as energy by utilising the relationship between the distance between neighbouring photon-assisted peaks (measured in $V_D$) and the microwave photon energy (see the supplementary information).

In addition, we observed the spin blockade phenomenon under certain double dot conditions. It is well known that weak leakage current flowing in spin blockade conditions indicates the occurrence of spin flip events in dots[33]. Under the electron spin resonance (ESR) condition of a single electron in a dot, the spin blockade is lifted and the leakage current increases when one of the spins is flipped by resonant microwaves in a static magnetic field[2, 6, 33]. As shown in Fig. 3(c), $I_S$ increases during ESR. Similar ESR responses are observable in the $(V_G, V_D)$ region enclosed by the dotted line in Fig. 3(a), strongly indicating that the spin blockade occurred in that region. Furthermore, the ESR response continues up to 12 K (Fig. 3(d)). Notably, the observable temperature is limited by a higher lying excited state of the satellite dot. In addition, the observable temperature corresponds to an energy much higher than the Zeeman energy of the spin, which can be advantageous in read-out using spin blockades, whereas single-shot read-out with spin-to-charge conversion requires a lower temperature than that corresponding to the Zeeman energy[34].

The amount of leakage current in the spin blockade region limits the mean stay time of an electron in the dot. In the experiments, we observed a leakage current of approximately 10 pA, which corresponds to a mean stay time of approximately 10 ns. This time is consistent with the inverse line width of the ESR of approximately 0.1 GHz at low microwave power.

The *g*-factor of the spin resonance shown in Figs. 3(c) and 3(d), which was obtained by applying a static magnetic field perpendicular to the substrate, was estimated to be 2.0. The *g*-factor changed slightly (by less than 2%) in the other spin blockade region (see the supplementary information). Note that these *g*-factors depend on the direction of the applied magnetic field and the resonance having a *g*-factor of 2.0 exhibited a *g*-factor of 2.3 when the magnetic field was applied parallel to the source-to-drain current flow (see the supplementary information). We note that the observed g-factor of 2.3 (much greater than 2) is largely different from that of the Boron impurity ($g = 1.1$)[35]. For Al-N impurity pairs, the large anisotropy of Zeeman splitting and *g*-factor of bound excitons ($g = 1.6$ for electron and $g = 1.1$ for hole) have been studied by magneto-photoluminescence[36]. However, the *g*-factor of the ground state probed by single electron tunnelling (accompanied by change in the electron/hole number) of Al-N will generally be different from the exciton *g*-factor. Thus, it has not been adequately proven that the observed *g*-factor is due to Al-N and this will be investigated in future research.

Considering the abovementioned results, the model of electron transport in a double dot shown in Fig. 3(e) was obtained. The spin blockade occurred in the double dot with deep and shallow impurities. It is supposed that the deep impurity was donor-like with a total spin of zero in the electron-poor state (electron number $N_e = 0$) and 1/2 in the electron-rich state ($N_e = 1$) and that the shallow impurity was acceptor-like with a total spin of 1/2 in the electron-poor state ($N_e = 1$, equivalent to the hole-rich hole number $N_h = 1$) and zero in the electron-rich state ($N_e = 2$, or $N_h = 0$). The opposite case, i.e., the combination of an acceptor-like deep impurity and a donor-like shallow impurity, is also possible.

**Coherent spin manipulation**

This section discusses the coherent spin operation of another double-dot-like device (device C). Figure 4(a) depicts the ESR response under spin blockade conditions, as in the case of device B. The background current is lower in this case than it was for device B, which indicates that a longer mean stay time of the spin in the dot was realised. Notably, extra current other than the dot-related ESR background current flowed in the device because of the wide gate width, causing the effective spin blockade leakage to be smaller than that observed and resulting in a longer effective mean stay time (see the supplementary information). The ESR peak width was determined to be 4 MHz for low microwave power, which is limited by the mean stay time of the spin blockade, rather than the width $(T_2^*)^{-1} \sim 1$ MHz being limited by the nuclear spins in natural Si[1, 3, 7, 8, 9, 10].

Pulse-modulated microwave irradiation with the resonant frequency coherently drove the spin, and clear Rabi oscillations with increasing pulse length are observable in Fig. 4(b). The oscillation amplitude (~0.1 pA) is consistent with the pulse repetition rate, or duration (1 μs). The frequency of the observed Rabi oscillations varies linearly with the square root of the microwave amplitude (Fig. 4(c)). During the off-period of the pulse train, the target dot is set at one of the spin triplet states due to the spin blockade; then, it becomes the initial state. The microwave driving changed the triplet state to the superposition of a singlet and triplets; thus, the singlet component could be read out as $I_D$. Figure 4(d) presents maps of the frequency detuning of the Rabi oscillations, displaying patterns well known to be evidence of qubit operation. Qubit operation is clearly observable up to 5 K and remains detectable at 10 K. The operation temperature of the TFET-based spin qubit is about two orders of magnitude higher than that of the other Si spin qubits. It should be noted that spin manipulation and read-out were realised at the same voltage due to strong electron confinement. In previous works with weak electron confinement[6, 33], spin manipulation required a voltage different from that used for read-out because strong microwave pulses collapse the electronic states of dots. Reduced visibility of the Rabi oscillation at 10 K is due to decrease in the Rabi amplitudes and increase in the noise level. Similar trend is seen in Fig. 3(d), where the decrease in the ESR peak and increase in the background noise eventually mask the ESR response at higher temperature. The $T_1$ value of the spin cannot be measured with our measurement scheme.

**Discussion**

In this work, we characterised 41 devices having channel lengths of 60, 70, and 80 nm. Among them, 37 devices exhibited single- or multiple-quantum-dot transport, all of which had high single-electron charging energies, as discussed above (see also the supplementary information). In terms of high-temperature operation, three devices (including devices A and B) exhibited single-electron transport at room temperature. In most cases, the gate leakage current hampered the observation of the single-electron transport. Therefore, the gate stack should be improved to realise significantly higher single-electron transport yields at room temperature.

Although the series of data clearly show the existence of the deep and shallow levels, the actual species of impurities are not identified yet. However, it is certain that the Al–N implantation caused the deep levels to form. To identify them, further investigation is required to consider the Al–N pair, implantation defects, and their complexes.

We discuss here that the deep impurity qubit can be compatible with both integrability and high-fidelity control, which are the necessary conditions for building a silicon quantum computer. A conventional qubit-qubit coupling method controls direct exchange interaction between two electron-spins, and thus the weakly localised electron wavefunction is preferred, but this leads to low-temperature operation due to weak confinement. However, schemes for long distance qubit-qubit coupling such as spin chains[37, 38], microwave cavity[39, 40] and dipole-dipole interactions[41] have been proposed. These are applicable for strongly localised electrons of the deep impurities, and enable high temperature qubit-qubit coupling. The employed qubit-readout method by the spin-blocked drain current is essentially a time ensemble measurement of the electron spin qubit, and cannot provide the high-fidelity non-demolition measurement required for a realistic quantum computer. However, if the deep impurity atom has a non-zero nuclear spin and it couples with the electron spin qubit by hyperfine interaction, this nuclear spin is regarded as a qubit, and can be read out with high fidelity and as a non-demolition measurement. There are proposals for such readout of nuclear spins in shallow or deep impurities[42, 43].

Improvement of operating temperature is recognised as an important development factor, even for architecture design, assuming implementation with conventional silicon quantum bits. It is believed to be difficult to realise the cooperation of qubits and classical CMOS circuits at such low temperatures because CMOS circuits introduce heat and electrical noise that hamper qubit operation. Therefore, at a minimum, qubit operation at 1–4 K is desirable[44].

In the future, we aim to realise spin blockade devices operating at even higher temperatures and with low variability sufficient for integration. A spin blockade requires a pair of donor- and acceptor-like levels, as discussed above. One level in the pairs was a shallow level in devices B and C in this work. Pairs of deep levels are expected to enable operation at higher temperatures.

In terms of variability, control of the impurity position is crucial, for which single-I/I technology can be employed[45]. Together with the realisation of such techniques, the Si-TFET-based spin qubits open the door to the age of high-temperature quantum technology on silicon technology.

**Materials and Methods**

In this study, we fabricated the TFETs using the 100-mm fabrication facilities at AIST. Several types of TFETs, such as those with different gate lengths and widths, were simultaneously fabricated on each wafer. The gate length varied from 10 μm to 60 nm. The long-channel TFETs with gate lengths longer than 100 nm were successfully operated as conventional TFETs. All of the transistors were operational; also, their variations were suppressed effectively[46]. On the other hand, the short-channel TFETs with gate lengths shorter than 90 nm exhibited a short-channel effect in which the OFF current increased because the drain bias had a stronger effect on the channel potential. The short-channel TFETs with gate lengths shorter than 80 nm operated as quantum transport devices, as reported in this manuscript.

**Conclusion**

We proposed and demonstrated quantum dots and spin qubits based on Si TFETs with deep impurities. The quantum dots operated at 300 K, whereas the spin qubits operated at temperatures up to 10 K, which was confirmed by spin manipulation with Rabi oscillation. Furthermore, the quantum dots and spin qubits effectively utilised deep impurity levels in Si. Future improvements will enable the development of high-temperature quantum electronic devices on silicon electronics.

**Data Availability:**

The datasets generated during and/or analysed during the current study are available from the corresponding author on reasonable request.

**Acknowledgements**

**General**: The measurements in this work were technically supported in part by the NIMS Nanofabrication Platform of the Nanotechnology Platform Project and WPI Initiative on Materials Nanoarchitectonics of MEXT, Japan. The authors thank Dr. Iizuka of NIMS, Japan, and Prof. Nakayama of Chiba University, Japan, for their valuable discussions regarding the theoretical views of deep levels. The authors thank Dr. Ota, Dr. Migita, Dr. Morita, Dr. Fukuda, Dr. Mizubayashi, Dr. Matsukawa, Dr. Masahara, and Dr. Yasuda of AIST, Japan, for their assistance with the device fabrication and characterisation. The authors thank Dr. Ishibashi, RIKEN, Japan; Prof. Eto, Keio University, Japan; Dr. Tanamoto, Toshiba Corporation, Japan; and Dr. Kawabata, AIST, Japan, for their valuable comments on this paper.

**Funding:** This work was financially supported by JSPS KAKENHI grants 15H04000, 15H05526, 16H03900, and 17H01276; and the JSPS FIRST Program initiated by CSTP, Japan.

**Author contributions:** T.M. designed and fabricated the devices. K.O. and S.M. measured and analysed the electronic properties of the devices. K.O., T.M., and S.M. wrote the manuscript.

**Competing interests:** The authors declare no competing interests.

**Figures**

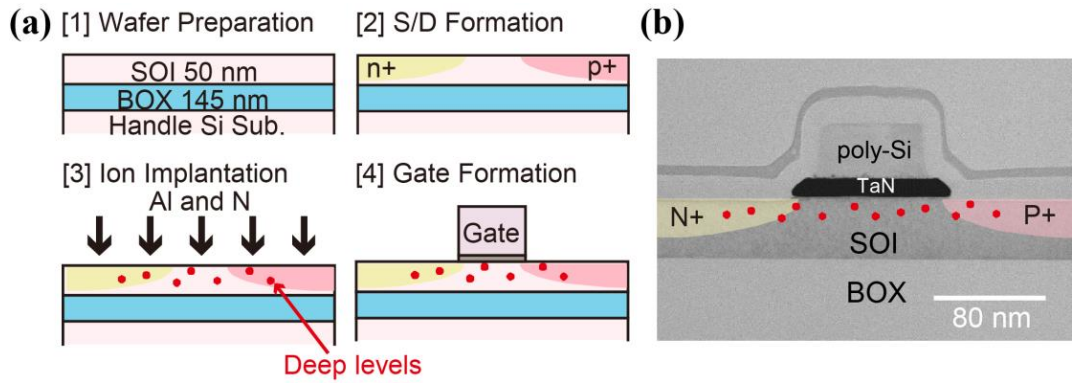

**Fig. 1. Device fabrication.** (**a**) Schematic cross-sections of device fabrication method. Details are provided in the supplementary information. (**b**) Typical transmission electron microscope image. BOX: buried oxide.

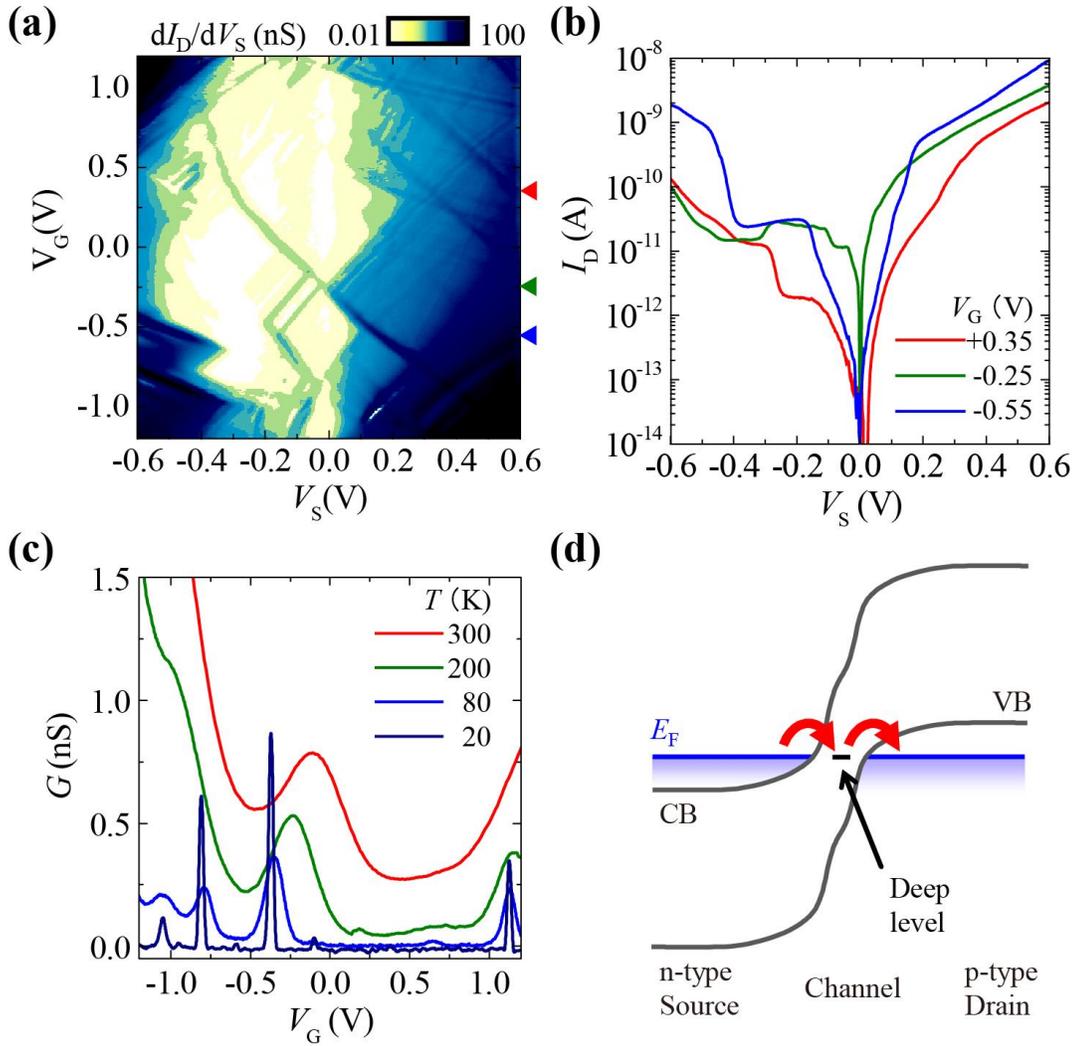

**Fig. 2. Characteristics of device A, the Al–N-implanted TFET with a channel length of 60 nm.** (**a**) Log-scale $dI_D/dV_S$ map at 40 K. The Coulomb diamonds exhibit the maximum widths at $V_G = 0.35$ V and $-0.55$ V, which correspond to single-electron charging energies of 0.3 eV and 0.1 eV, respectively. The high currents near all four corners are also observed for conventional TFETs and originate from band-to-band tunnelling for positive $V_S$ and diffusion current for negative $V_S$ (see also the supplementary information). (**b**) Log-scale $I_D$–$V_S$ at various $V_G$, which are marked using triangles of the same colours as in (a). Coulomb staircases are observable (particularly in the negative $V_S$). In addition, a negative Coulomb staircase is observable at $V_S = -0.3$ V in the $V_G = -0.25$ V curve (see the supplementary information). (**c**) $V_G$ dependence of $G$ at various temperatures. With increasing temperature, the conductance peak is positively shifted, which is supposed to be due to the MOS capacitance change caused by thermally activated carriers in the channel (see the supplementary information). (**d**) Schematic of the band diagram of a TFET with a deep impurity, along with the MOS interface, where CB, VB, and $E_F$ denote the conduction band, valence band, and Fermi energy, respectively. Single-dot-like electron transport occurred through the deep impurity level.

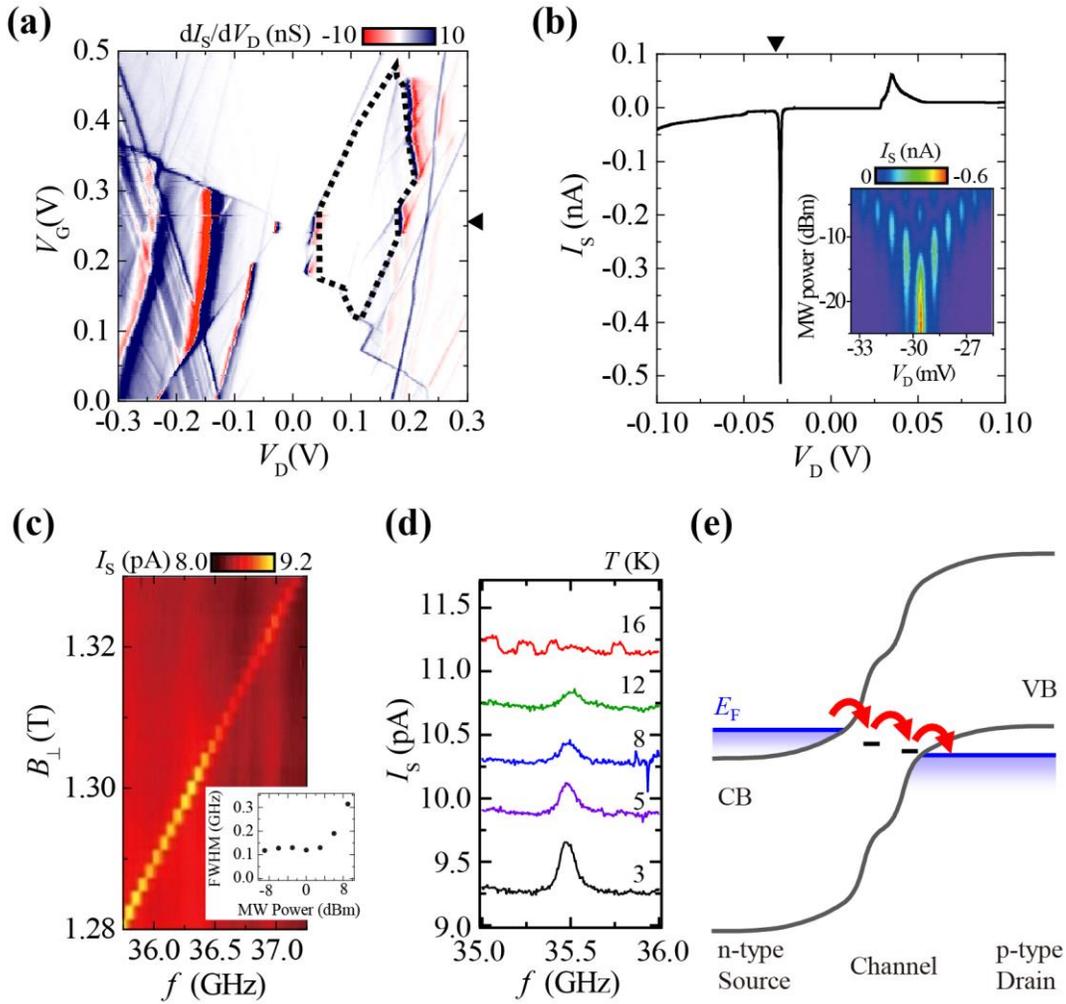

**Fig. 3. Characteristics of device B, the Al–N-implanted TFET with a channel length of 70 nm. All of the measurements were conducted at 1.5 K unless otherwise noted.** (**a**) Linear-scale $dI_S/dV_D$ map. (**b**) $I_S$–$V_D$ at $V_G = 0.253$ V (marked by a triangle in (a)). The inset depicts the variation in the current during microwave irradiation with $V_D$ (near the sharp peak at −0.03 V, marked by the triangle in the main graph) and microwave power at a constant frequency of 32.7 GHz. (**c**) $I_s$ at $(V_D, V_G) = (0.055$ V, $0.253$ V) as a function of magnetic field $B_\perp$ (applied perpendicular to the substrate) and microwave frequency with a constant power $P$ of 0 dBm (at the output of the microwave power source), showing ESR. The ESR response was measured for every other $(V_D, V_G)$ set in (a), with 5 mV intervals for each voltage. The similar ESR is evident in the enclosed area marked by the dotted line in (a). Inset shows microwave power dependence of the ESR linewidth. (**d**) Temperature dependence of the ESR peak observed at $(V_D, V_G) = (0.06$ V, $0.25$ V) with $B_\perp = 1.255$ T and $P = 3$ dBm. Note that the as-measured curves are shown without any artificial base current offset; thus, the higher the temperature, the higher the base current. (**e**) Schematic band diagram for double-dot-like transport with a deep impurity and an acceptor-like shallow impurity.

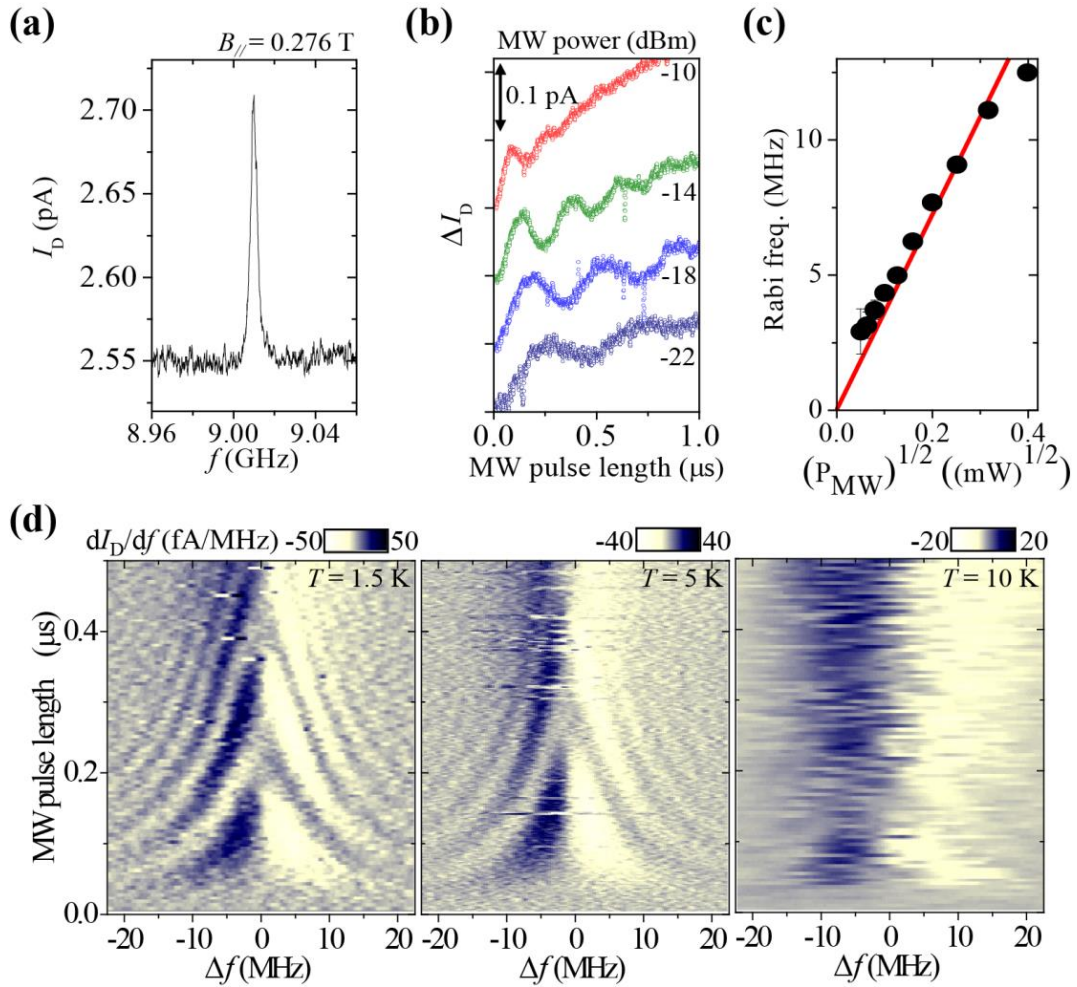

**Fig. 4. Characteristics of device C, the Al–N-implanted TFET with a channel length of 80 nm. All of the measurements were conducted at 1.5 K unless otherwise noted.** (**a**) ESR with a width of 4 MHz observed at ($V_D$, $V_G$) = (0.33 V, −0.36 V) in a magnetic field $B_{//}$ = 0.276 T (applied parallel to the source-to-drain current direction) and with $P$ = −18 dBm. The ESR is similar to those obtained using other ($V_D$, $V_G$) sets, as in the case of device B. The g-factor was estimated to be 2.3, and consistent with device B for $B_{//}$. A weak ESR with a g-factor of 2.7 was also observed (see the supplementary information). (**b**) Steady-state change in the drain current $\Delta I_D$ as a function of the pulse length of the pulse-modulated microwaves (simple pulse train with fixed pulse repetition of 1 μs). The microwave frequency was fixed at that necessary for ESR (~9.01 GHz) with 2-μs pulse repetition. The plots are offset for clarity. The background current increase with increasing pulse length is due to the static shift in the effective $V_G$ caused by the microwave. (**c**) Oscillation frequency dependence on $P$. (**d**) Oscillation dependence on microwave detuning $\Delta f$ at 1.5, 5, and 10 K. The microwave pulse repetition was 1 μs for all of the measurements, and $P$ was −11, −13, and −10 dBm for the measurements at 1.5, 5, and 10 K, respectively. The characteristic Rabi amplitudes (at π-pulse) are 0.08 pA, 0.08 pA, and 0.04 pA and noise values are 0.01 pA, 0.01 pA, and 0.02 pA at 1.5 K, 5 K, and 10 K, respectively.

**Supplementary Information for High-temperature operation of a silicon qubit**

Keiji Ono, Takahiro Mori, and Satoshi Moriyama

**Fabrication of TFETs**

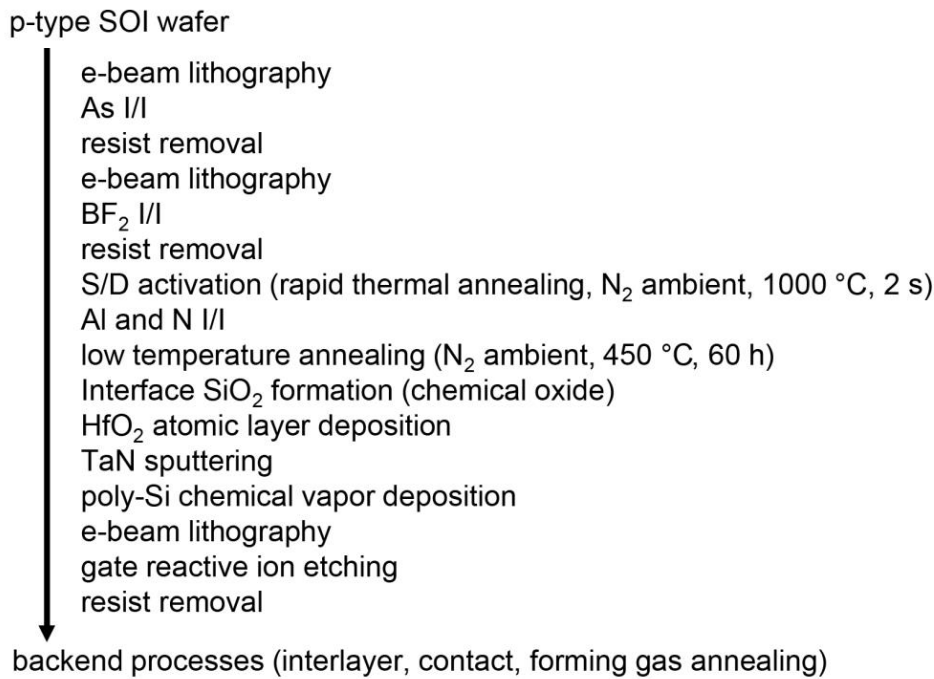

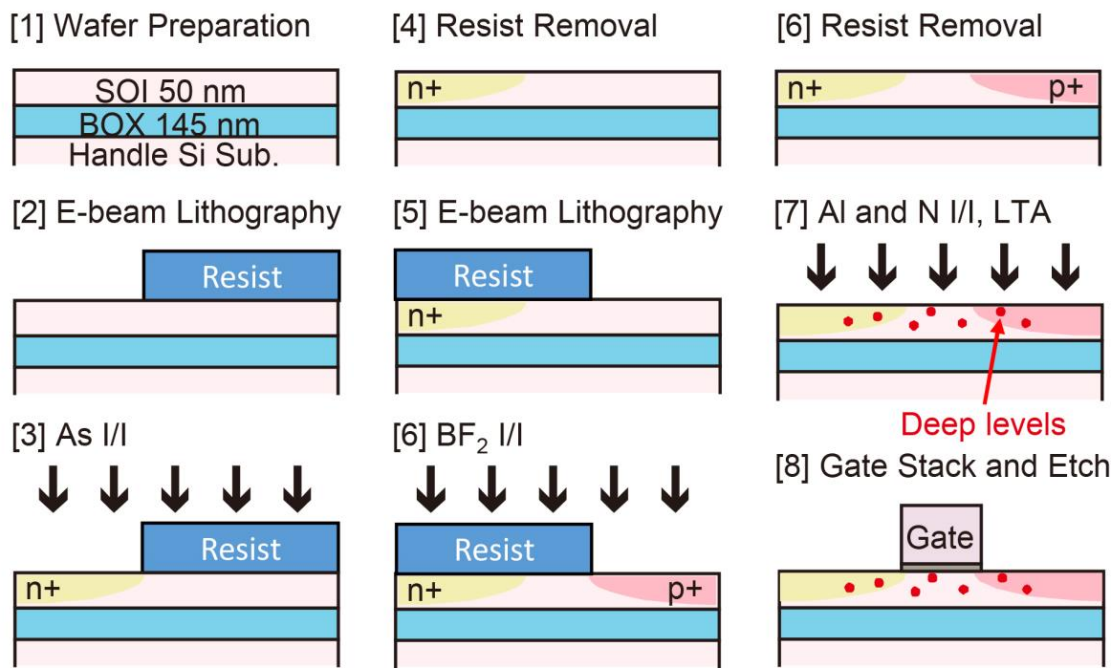

**Fig. S1. Fabrication of TFETs.** (**a**) Device fabrication process. (**b**) Schematic cross-sections of some of the steps in the process.

TFETs are fabricated via a process similar to that used to produce MOSFETs, which are the bases of conventional LSIs. TFETs are gated p-i-n diodes, in which the tunnelling current flowing through a pn junction is regulated at the source-side edge of the gate using electrostatic gate control. For example, in P-type TFETs, the source is n-type, whereas the drain is p-type. It should be noted that the source and drain definitions used in this paper are the same as those in the conventional P-type TFET case, although quantum transport of TFETs is discussed in this report.

We fabricated the TFETs on SOI wafers (Fig. S1). The SOI thickness was approximately 50 nm, and the BOX thickness was 145 nm. The top wafer surface was (100), and the current flow direction was <110>. First, I/I was employed to produce the source and drain. The source was formed by BF2 I/I with energy of 5 keV and a dose of $2 \times 10^{15}$ cm$^{-2}$. The drain was formed by As I/I with the same energy and dose. Rapid thermal annealing was performed at 1000 °C for 1 s to activate the source and drain. Then, we introduced isoelectronic traps (IETs) into the active region by I/I of Al and N with energy of 15 keV and a dose of $2 \times 10^{13}$ cm$^{-2}$. Low-temperature annealing was conducted at 450 °C for 60 h to activate the Al–N IETs. Following the introduction of the IETs, the deep levels we utilised in this work were formed in the channel. Finally, the gate was produced using high-k/metal gate technology. The interfacial SiO$_2$ was chemically formed with a thickness of approximately 1 nm, and HfO$_2$ (2.4 nm) was deposited onto it by performing atomic layer deposition at 250 °C. A TaN gate was formed by sputtering with a thickness of 10 nm, and a 50-nm doped-poly-Si cap was deposited by conducting chemical vapour deposition. This gate structure is also referred to as a metal-inserted-poly-Si gate. The equivalent oxide thickness of the gate insulator was estimated to be approximately 1.5 nm based on capacitance-voltage measurements. The IET technology utilised in this work has been proposed to improve the active performance of Si-based TFETs by enhancing the ON current. It should also be noted that the Al–N IET dose was 10 times higher than that in previous reports on conventional TFETs[20, 21, 22].

In this study, we fabricated the TFETs using the 100-mm fabrication facilities at AIST. Many kinds of TFETs, such as ones with different gate lengths and widths, were simultaneously fabricated on each wafer. The long-channel TFETs with gate lengths longer than 100 nm were successfully operated as conventional TFETs. All of the transistors were operational; also, their variations were suppressed effectively[46]. On the other hand, the short-channel TFETs with gate lengths shorter than 90 nm exhibited a short-channel effect in which the OFF current increased because the drain bias more strongly affected the channel potential. The short-channel TFETs operated as quantum transport devices, as reported in this manuscript.

**Differential conductance maps of TFETs (conventional short-channel and Al–N-implanted long-channel)**

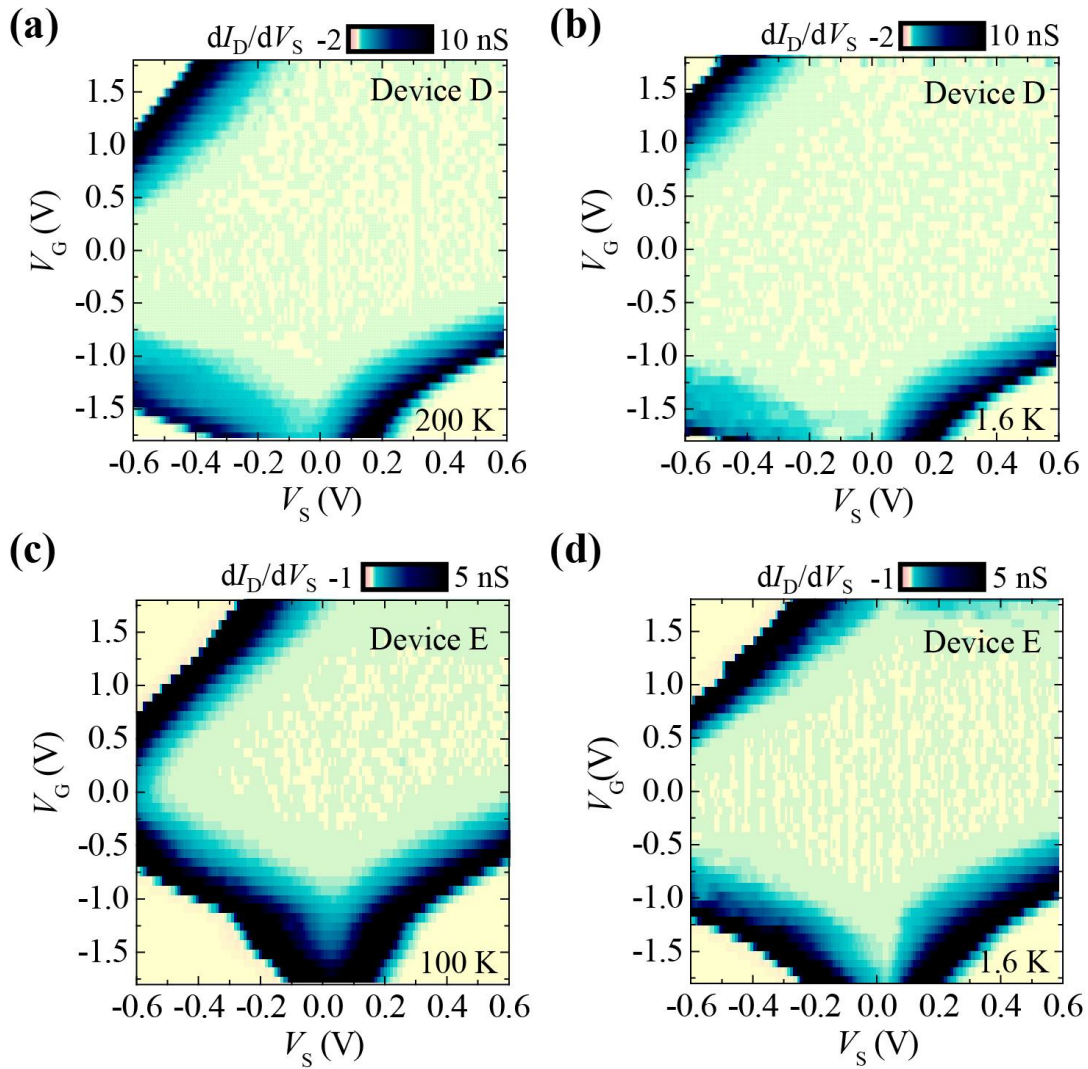

**Fig. S2. Differential conductance maps of TFETs.** (**a**), (**b**) $dI_D/dV_S$ intensity maps obtained at 200 K and 1.6 K, respectively, for device D, which was the TFET without Al–N implantation and with a channel length of 60 nm. (**c**), (**d**) $dI_D/dV_S$ intensity maps obtained at 100 K and 1.6 K, respectively, for device E, which was the Al–N-implanted TFET with a channel length of 100 nm.

The TFETs without Al–N implantation did not produce Coulomb diamonds, although they had short channels, as shown in Fig. S2(a) and (b) for the device with a channel length of 60 nm as an example. The Al–N-implanted TFETs with long channels also did not produce Coulomb diamonds, as shown in Fig. S2(c) and (d). The increase in $dI_D/dV_S$ in all four corners of the maps are in accordance with those observed in conventional TFET operation. The increase in the lower-right corners originated from band-to-band tunnelling between the channel (in which holes were generated when negative gate bias was present) and n-type source. Band-to-band tunnelling is also observable in the upper-right corner of Fig. S2(d), where tunnelling occurs between the channel (in which electrons were generated) and the p-type drain. The increase in the upper- and lower-left corners originated from the diffusion current, which corresponds to the forward-bias characteristics of conventional pn diodes. These features are mostly independent of temperature.

**Temperature dependence and negative Coulomb staircase of device A**

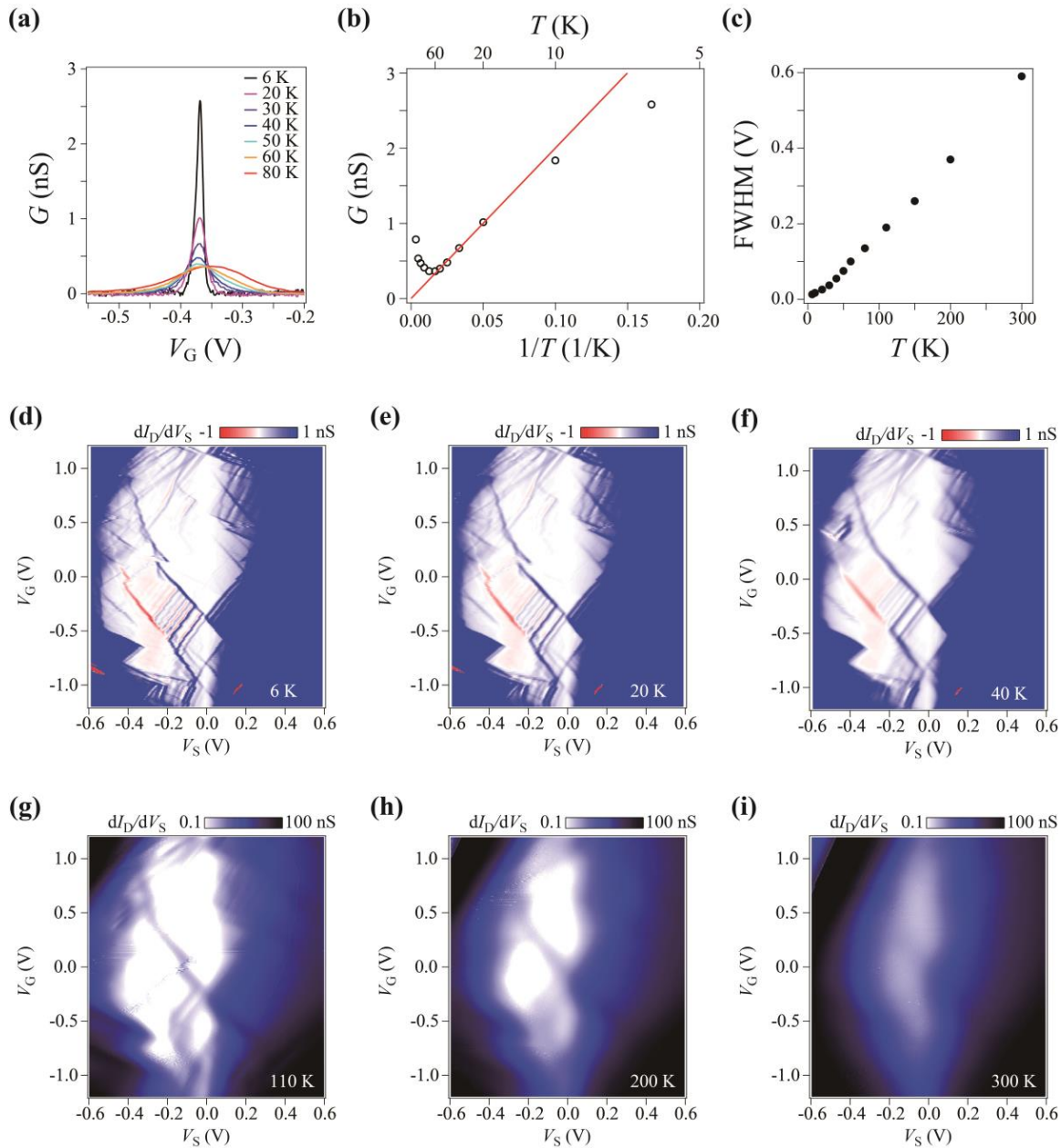

**Fig. S3. Temperature dependence and negative Coulomb staircase of device A.** (**a**) Detailed temperature dependence of the conductance peak in Fig. 2(c) for the lower temperature range. The peak was observable at temperatures up to 300 K. (**b, c**) Temperature dependences of the peak height (b) and width (c) of the conductance peak. (**d, f**) Linear-scale $dI_D/dV_S$ intensity maps measured at 6 K (**d**), 20 K (**e**), and 40 K (**f**). (**g, f**) Log-scale $dI_D/dV_s$ intensity maps measured at 110 K (**g**), 200 K (**h**), and 300 K (**i**).

Figure S3(a–c) present the detailed temperature dependence of the conductance peak in device A, which was observable at temperatures up to 300 K. As indicated by the red line in Fig. S3(b), for 20 K ≤ $T$ < 60 K, resonant tunnelling through the dot is dominant, and the conductance peak shapes agree closely with those obtained from the theoretical expression for the quantum Coulomb blockade[47], where the peak height is proportional to $1/T$. At lower temperatures ($T$ < 20 K), the conductance peak height appears to approach saturation. For 60 K ≤ $T$ < 110 K, the conductance peak height is independent of temperature. The transition from quantum to classical Coulomb blockade occurs in this range. At temperatures greater than 110 K, the conductance peak height increases. This is not in accordance with classical Coulomb blockade theory, probably because of superposition of the tails of neighbouring peaks, current leakage and generation, and parallel conduction paths in the channel. The transition temperature around 60 K is in agreement with the level spacing between the first excited and ground states, $\Delta E = 20$ meV, as is observable in Fig. S3(d–f). The conductance peak width corresponds to ~4 $k_B T$ for a quantum Coulomb blockade; in this case, 4 $k_B T$ is nearly 20 meV when $T = 60$ K. The FWHM of the conductance peak monotonically increases with increasing temperature, mostly following the theory. The charging energy of 0.1 eV estimated from the temperature dependence of the Coulomb peak width is consistent with that estimated from the Coulomb diamonds.

Figure S3(d–f) depict the linear-scale intensity maps of $dI_D/dV_S$ at low temperatures. Negative differential conductance is observable in the red regions with $V_S$ ranging approximately from −0.2 V to −0.4 V and $V_G$ approximately from 0 V to −0.6 V. The thick red lines slanting down to the right correspond to the negative staircase in Fig. 2(b). The thin red lines slanting down to the left are along the edges of the Coulomb diamonds and parallel to the excited states. In previous studies, it was theoretically predicted that negative Coulomb staircases should be observed in single-molecule devices in which single-electron transport is coupled to the vibration mode of the molecule[48, 49]. Therefore, it is suggested that the quantum dots in our devices were atomically small, so molecule-like vibration modes existed. We suppose that the vibrations originated from local phonon modes of the deep impurities.

**Temperature-dependent characteristics of device B**

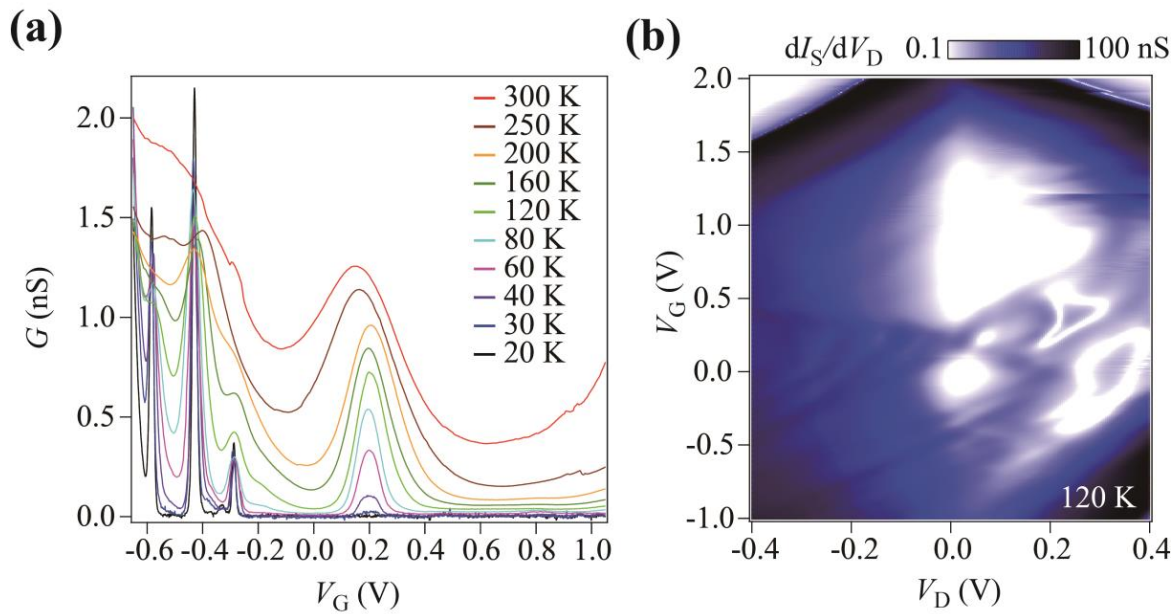

**Fig. S4. Temperature-dependent characteristics of device B.** (**a**) Temperature dependence of the zero-bias conductance in device B. (**b**) Log-scale d$I_S$/d$V_D$ intensity map measured at 120 K.

Figure S4(A) shows the temperature dependence of the zero-bias conductance in device B. The peak at $V_G \sim 0.2$ V is observable at temperatures above 40 K; however, it is suppressed below 30 K. The peak position is in the double-quantum-dot transport region (Fig. 3(a)). Therefore, the suppression at low temperatures is due to the double dot formation, and the peak appears at high temperatures due to the transition from double to single dot because the weakly confined dot vanished. The peak is observable up to 300 K, which indicates that one of the dots had strong confinement energy originating from a deep level. Figure S4(b) shows the log-scale d$I_S$/d$V_D$ intensity map measured at 120 K. Large, closed Coulomb diamonds are observable, with maximum widths at $V_G = 0$ V and 0.9 V corresponding to large single-electron charging energies of 0.2 eV and 0.3 eV, respectively. This strongly confined single dot behaviour can be described as discussed above.

# Landau-Zener-Stuckelberg-Majorana interference in device B

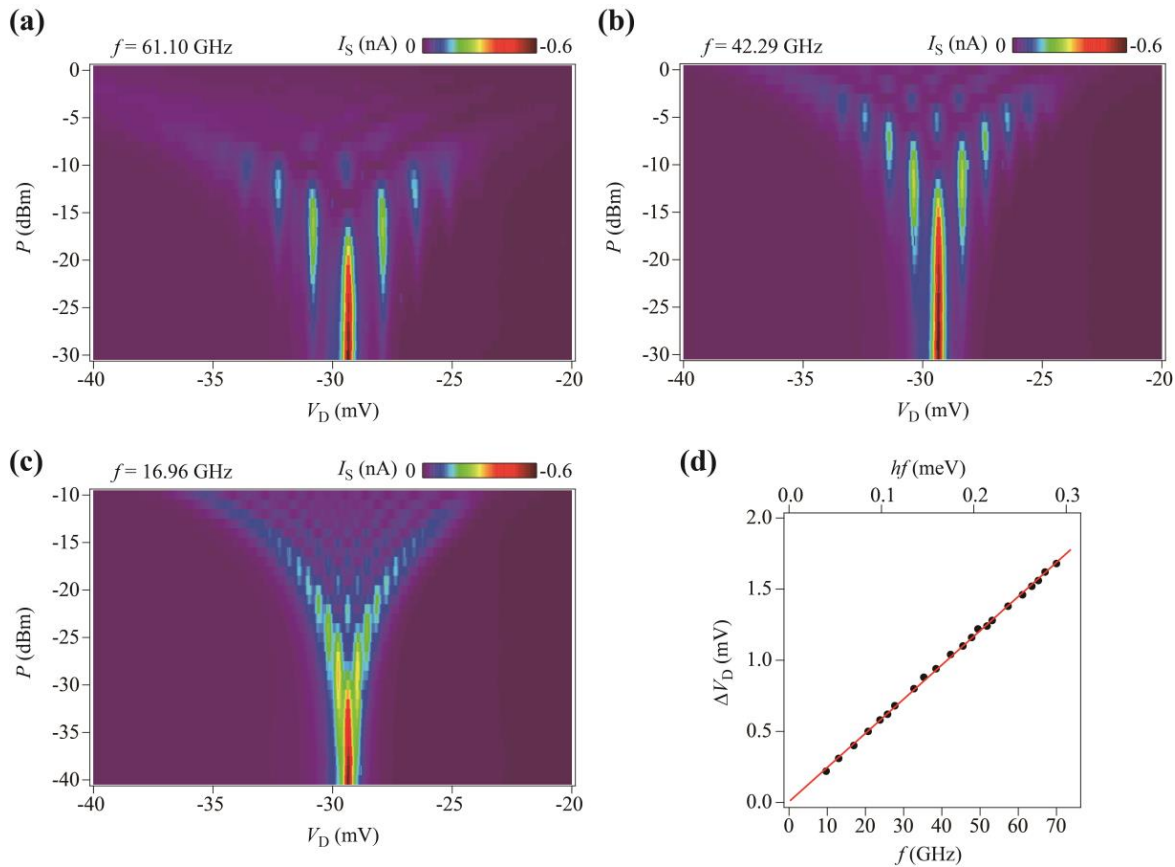

**Fig. S5. Landau-Zener-Stuckelberg-Majorana interference in device B.** (**a–c**) Colour intensity maps of $I_S$ near the sharp peak at $V_D = -0.03$ V (Fig. 3(b)) measured at 1.5 K as functions of $V_D$ and $P$ at $V_G = 0.255$ V with constant microwave frequencies of 61.10 GHz (a), 42.29 GHz (b), and 16.96 GHz (c). (**d**) Voltage difference $\Delta V_D$ between the first satellite peaks and main peak as a function of the applied microwave frequency.

Figure S5(a–c) present the dependence of the sharp $I_S$ peak at $V_D = -0.03$ V (Fig. 3(b)) on $P$. We observed Landau-Zener-Stuckelberg-Majorana interference at frequencies ranging from 10 GHz to 70 GHz. The dependence of the peak height on $P$ agrees with the theoretical square Bessel function expression[35]. Landau-Zener-Stuckelberg-Majorana interference is observable even if the energy of one photon $hf$ is smaller than the measurement temperature (1.5 K). $\Delta V_D$ depends linearly on the frequency between 10 GHz and 70 GHz, as shown in Fig. S5(d). Using the slope of the line, we estimated the conversion factor between the drain voltage and energy in the dot as 0.17 meV/mV (17%). Using this factor, the $I_S$ peak width was converted into energy of 0.06 meV, which is less than that corresponding to the measurement temperature (0.12 meV, corresponding to 1.5 K). This relationship indicates that the electron transport is limited by the lifetime of the single-particle states in the quantum dots and is independent of temperature, and provides evidence of resonant tunnelling in the series-coupled double quantum dot.

# ESR in device B

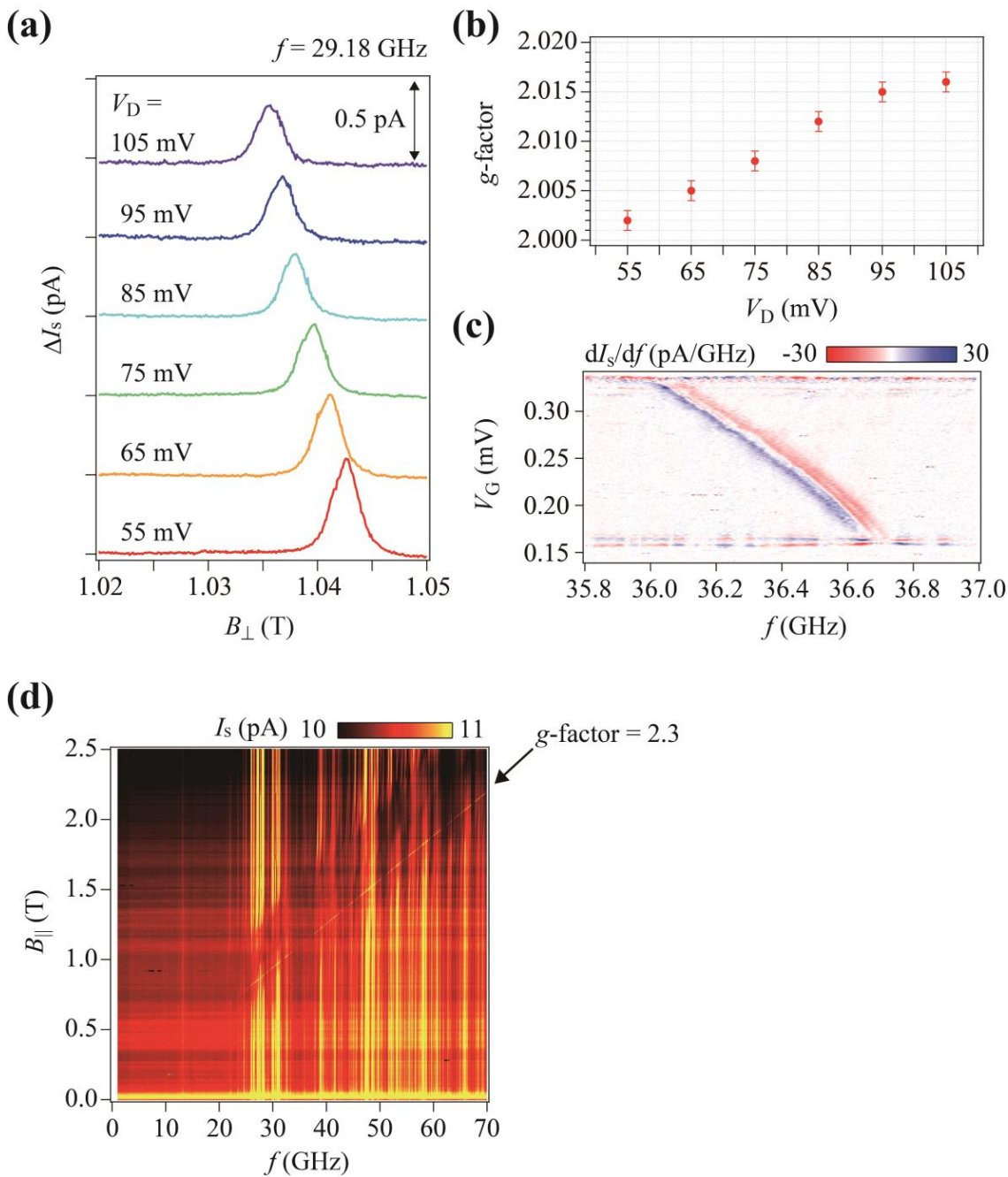

**Fig. S6.** (a) $V_D$ dependence of the ESR peak as a function of $B_\perp$ with $V_G = 0.253$ V, $f = 29.18$ GHz, and $P = 3$ dBm. Each curve is artificially shifted by 0.5 pA from the one below it for clarity. (b) $V_D$ dependence of the $g$-factor obtained from Fig. S6(a) and similar measurements with several frequencies and magnetic fields. We performed measurements with various $B_\perp$ and confirmed that the peak position vs. $B_\perp$ line passed through the origin. (c) $V_G$ dependence of the ESR peak ($dI_S/df$) as a function of microwave frequency with $P = 3$ dBm, $V_D = 0.075$ V, and $B_\perp = 1.3$ T. For $0.16$ V $< V_G < 0.33$ V, the ESR response is observable because a spin blockade occurred in this range (see also Fig. 3(a)). (d) $I_S$ intensity of device B at $(V_D, V_G) = (0.055$ V, $0.14$ V$)$ as a function of $B_\parallel$ and microwave frequency with $P = 3$ dBm and different cooldowns, measured at 1.5 K.

Figure S6(a) shows the $V_D$ dependence of the ESR peak as a function of $B_\perp$ in the spin blockade transport region (see also Fig. 3(a)). We performed similar measurements using several frequencies and magnetic fields, and the $g$-factors obtained from the dependence of the ESR peak position on $B_\perp$ for each $V_D$ are summarised in Fig. S6(b). Figure S6(c) depicts the $V_G$ dependence of $dI_S/df$ in the spin blockade transport region (see Fig. 2(a)). The change in the peak position is attributed to the change in the $g$-factor from 1.98 to 2.01. Specifically, in the spin blockade region, which was the area enclosed by the dotted line in Fig. 3(a), the $g$-factor changed by approximately 2%.

Figure S6(d) presents the ESR spectra of device B obtained with different $B_\parallel$ and different cooldowns. After several thermal cycles in which the temperature varied from 300 K to 1.5 K, the structure of the charge stability diagram hardly differs from that in Fig. 3(a), but the diagram is shifted towards negative gate voltages by about 0.1 V, probably because of the generation of fixed charge in the gate oxide due to the thermal cycling. Considering the voltage shift, we performed the same measurements as those used to obtain Fig. 3(c), except with a different magnetic field direction, and obtained a $g$-factor of 2.3.

## Coulomb diamond and ESR spectra of device C

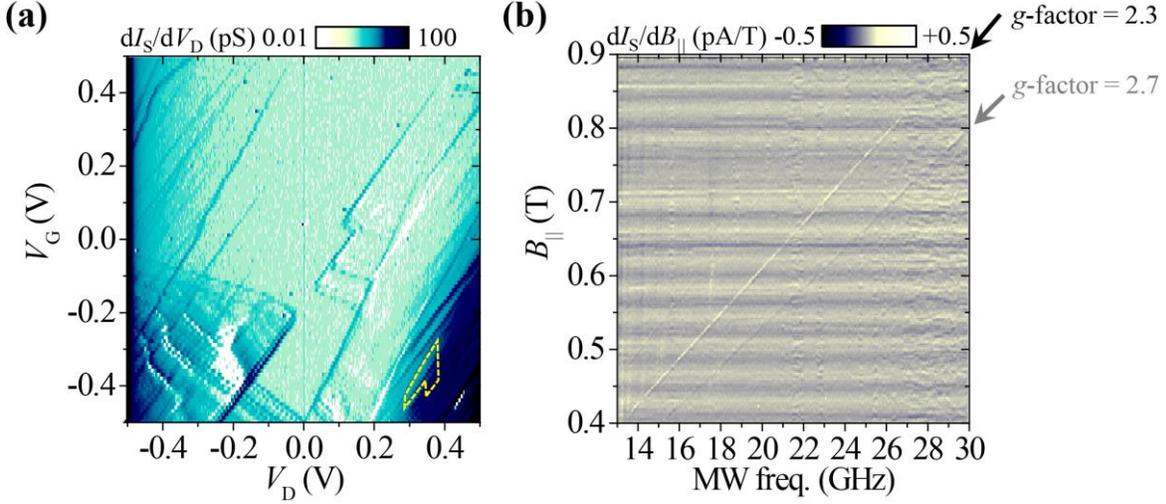

**Fig. S7. Coulomb diamond and ESR spectra of device C.** (**a**) d$I_S$/d$V_D$ intensity map obtained at 1.5 K, in which the spin blockade area is enclosed by a yellow dotted line near the lower-right corner. (**b**) d$I_D$/d$B_{//}$ intensity map measured at ($V_D$, $V_G$) = (0.33 V, −0.36 V) and 1.5 K as a function of $B_{||}$ and microwave frequency.

The Coulomb diamond and ESR spectra of device C that were measured at 1.5 K are presented in Fig. S7. The intensity of d$I_S$/d$V_D$ is weaker than it was for device B because the channel length of device C was longer than that of device B, which resulted in weaker tunnel coupling in device C. The open (unclosed) diamond at $V_G$ = −0.15 V suggests that multiple dots were formed in the channel. We measured the ESR responses for every other ($V_D$, $V_G$) set in 10 mV intervals in Fig. S7(a). ESR response similar to that in Fig. 4(a) occurred in the area enclosed by the yellow dotted line in Fig. S7(a). This area is smaller than the corresponding area for device B, and its shape is different from the conventional one, which suggests that complicated parallel conduction paths existed in the channel.

Figure S7(b) shows the ESR spectra measured in the spin blockade region. Two ESR lines with *g*-factors of 2.3 and 2.7 are observable. The peak intensity of the ESR with the *g*-factor of 2.7 is weak, so it is barely recognizable in the d$I_D$/d$B_{//}$ map. Thus, pulsed ESR could not be conducted with that ESR peak.

**Summary of other short-channel Al–N-implanted TFETs**

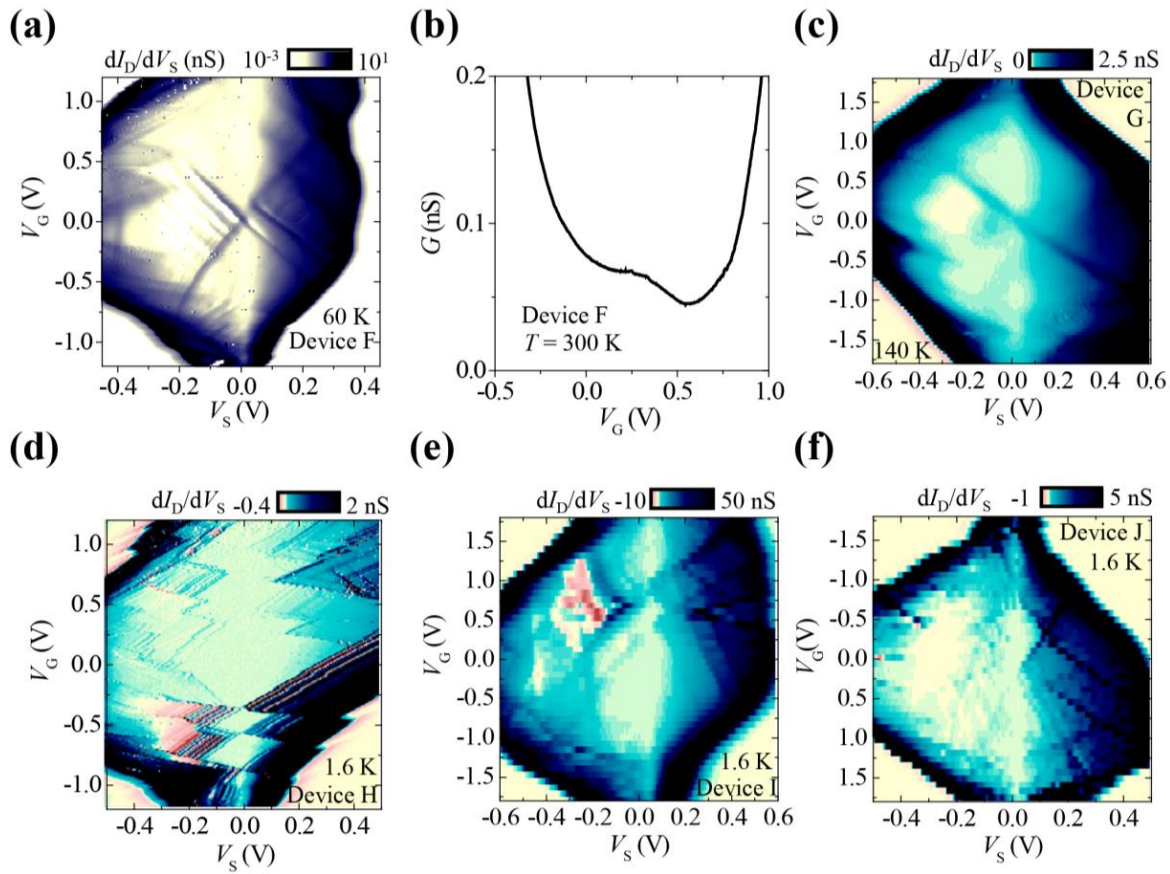

**Fig. S8. Characteristics of five Al–N-implanted short-channel TFETs.** (**a**) $dI_D/dV_S$ intensity map measured at 60 K for device F, which had a channel length of 70 nm. (**b**) $G$–$V_G$ curve measured at 300 K for device F. (**c**) $dI_D/dV_S$ intensity map measured at 140 K for device G, which had a channel length of 70 nm. (**d**–**f**) $dI_D/dV_S$ intensity maps measured at 1.6 K for device H, which had a channel length of 70 nm (**d**); device I, which had a channel length of 60 nm (**e**); and device J, which had a channel length of 60 nm (**f**).

As discussed in the main text, we characterised 41 devices with channel lengths of 60, 70, and 80 nm. Among them, 37 devices, all of which had high single-electron charging energies, exhibited single- or multiple-quantum-dot transport. Furthermore, three devices (devices A, B, and F) exhibited single-electron transport at room temperature. The characteristics of device F are depicted in Figs. S8(a) and S8(b). Device G (whose characteristics are illustrated in Fig. S8(c)) operated at temperatures up to 140 K, although the charging energy was as high as those of the three aforementioned devices were. The characteristics of three additional devices are provided in Fig. S8(d–f) as examples of multiple-dot-like transport at 1.6 K.